\documentclass[twocolumn, aps, showpacs]{revtex4}
\usepackage{amsmath}

\begin{document}

\title{Depinning transition at the upper critical dimension}
\author{Andrei A. Fedorenko and Semjon Stepanow}
\affiliation{Martin-Luther-Universit\"{a}t Halle-Wittenberg, Fachbereich Physik, D-06099
Halle, Germany}

\date{\today}

\begin{abstract}
We study the effect of quenched random field disorder on a driven elastic
interface close to the depinning transition at the upper critical dimension $%
d_{c}=4$ using the functional renormalization group. We have found that the
displacement correlation function behaves with distance $x$ as $(\ln
x\Lambda _{0})^{2/3}$ for large $x$. Slightly above the depinning transition
the force-velocity characteristics is described by the equation $\mathrm{v}%
\sim f\left\vert \ln f\right\vert ^{2/9}$, while the correlation length
behaves as $L_{\mathrm{v}}\sim f^{-1/2}\left\vert \ln f\right\vert ^{1/6}$,
where $f=F/F_{c}-1$ is the reduced driving force.
\end{abstract}

\maketitle

The driven viscous motion of an elastic interface in a medium with randomly
distributed pinning disorder has attracted considerable theoretical interest
during the last decade and is in a state of rapid development. The reason is
that, on one hand elastic interfaces in a disordered medium exhibit the rich
behavior of glassy systems and on the other hand it can serve as a model for
many experimental systems, such as domain walls in magnetically or
structurally ordered systems with impurities and interfaces between
immiscible fluids in porous media. Other closely related problems are the
motion of a vortex line in an impure superconductor and the motion of a
dislocation line in a solid [\onlinecite{fisher-phys-rep98}-%
\onlinecite{chauve}]. In all these systems the basic physical ingredients
are identical: the elastic forces tend to keep the interface flat, whereas
the disorder locally promote the wandering. In the dynamics, this interplay
between quenched disorder and elasticity leads to the complicated response
of the interface to an externally applied force. At zero temperature, a
driving force $F$ exceeding certain threshold value $F_{c}$ is required to
set the elastic interface in steady motion. The depinning transition can be
considered as a nonequilibrium dynamical critical phenomenon \cite%
{fisher-phys-rep98} where a system becomes extremely sensitive to small
perturbation. Recently, significant progress has been made in understanding
the depinning transition \cite{nstl92,nf93} (for recent studies see [%
\onlinecite{chaveetal-cond-mat}-\onlinecite{LeDousal-cond-mat02}]). It has
been shown that the functional renormalization group (FRG) gives an adequate
description of the critical behavior at the depinning transition if one
presumes to consider a singular renormalized random force correlator. The
scaling analysis shows that the disorder effects dominate over elasticity in
dimensions $d<4$, and therefore $d_{c}=4$ is the upper critical dimension of
the problem. Below $d_{c}$ the interface undergoes the depinning transition
at a critical driving force $F_{c}$ and slightly above the critical force $%
F_{c}$ the average velocity of the interface behaves as
\begin{equation}
\mathrm{v}\sim (F-F_{c})^{\theta },\ \ \ \ \ \ \ F>F_{c},  \label{v-F}
\end{equation}
where $\theta $ being the critical exponent. The roughness exponent
characterizing the width $w$ of the wandering interface at the depinning
transition is defined by
\begin{equation}
w\sim L^{\zeta },  \label{zeta}
\end{equation}
where $L$ is the linear size of the interface. The FRG analysis carried out
in Ref.~\cite{nstl92} enabled one to compute the critical force $F_{c}$, and
the critical exponents $\theta $ and $\zeta $ to order $\varepsilon =4-d$.
In the case of random force disorder it was found that $\theta
=1-\varepsilon /9+O(\varepsilon ^{2})$ and $\zeta =\varepsilon /3$ \cite%
{nstl92,nf93}. It was suggested in \cite{nf93} that at the depinning
transition the critical exponents for random bond and random field (RF)
disorder are the same. Initially, the result\ for the roughness exponent was
expected to be exact for all $d=1,2,3$ \cite{nf93}, but more recently, the
non-zero two-loop correction to $\zeta $ has been found \cite{chauve}.

The results of FRG analysis have been checked by intensive numerical studies
using both direct simulation [\onlinecite{ji-robbins92}-%
\onlinecite{rosso2002}] and different cellular automata models [%
\onlinecite{leschhorn93}-\onlinecite{rosso2001}], which are believed to
belong to the same universality class. The computed values of critical
exponents are in a good agreement with the predictions of FRG at least for $%
d=1,2,3$. In the numerical works \cite{rosso2002,roters2002} the depinning
transition was studied at the upper critical dimension. However, to our
knowledge, no explicit consideration of the depinning transition at the
upper critical dimension $d_{c}=4$ is available so far. Another motivation
to consider the depinning transition at the upper critical dimension is that
some experimental elastic systems, for example systems with dispersive
elastic constant such as moving geological faults arising from earthquakes
\cite{fisher-phys-rep98}, or systems with long-range Coulomb interaction,
have the upper critical dimension $d_{c}=3$ or $2$. One expect that these
systems may show a behavior which is similar to the behavior of a simple
model at $d_{c}=4$ \cite{chitra99}.

It is well-known \cite{zinn-justen} that at the upper critical dimension the
power laws modify to logarithmic corrections. While at the upper critical
dimension the one-loop RG consists in summing the main logarithms, the
two-loop RG takes into account the subdominant logarithms \cite%
{bogoliubov-shirkov-book}. Due to the fact that close to the depinning
transition the main logarithms are leading, the results of the one-loop RG
treatment are expected to be exact at the upper critical dimension for $%
F\rightarrow F_{c}$.

In this Brief Report we consider the motion of an elastic interface in a
disordered medium and our main purpose is to describe the critical dynamics
near depinning threshold for $d=d_{c}$ by using FRG method to one-loop
order. The motion of a $d$-dimensional interface obeys the equation
\begin{equation}
\lambda \frac{\partial z(x,t)}{\partial t}=\gamma \nabla ^{2}z+F+g(x,z),
\label{motion}
\end{equation}%
where $\lambda $ is the friction coefficient (or the inverse mobility), $%
\gamma $ is the stiffness constant, and $F$ is the driving force density.
The quenched random force $g(x,z)$ is assumed to be Gaussian distributed
with the zero mean and the correlator
\begin{equation}
\left\langle g(x,z)g(x^{\prime },z^{\prime })\right\rangle =\delta
^{(d)}(x-x^{\prime })\Delta (z-z^{\prime }).  \label{correl}
\end{equation}%
To make this model well-defined one has to introduce the cutoff $\Lambda
_{0}^{-1}$ in the $\delta ^{d}(x)$ function at scales of order the impurity
separation or other microscopic scales. We restrict our consideration to the
case of random field disorder when the correlator $\Delta (z)=\Delta (-z)$
is a monotonically decreasing function of $z$ for $z>0$ and decays rapidly
to zero over a finite distance.

In Ref.~\cite{nstl92} the RG analysis of the model (\ref{motion}), (\ref%
{correl}) was carried out using the technique of path integrals in the
one-loop approximation. After integrating out fluctuations in the momentum
shell $\Lambda _{l}<\left\vert k\right\vert <\Lambda _{0}$, the following RG
flow equations have been obtained
\begin{eqnarray}
&&\frac{d\ln \lambda }{dl}=-\frac{K_{d}}{\gamma ^{2}\Lambda
_{l}^{\varepsilon }}\int\limits_{0}^{\infty }dt\,\ te^{-t}\Delta ^{\prime
\prime }(\widetilde{\mathrm{v}}t),  \label{lambda} \\
&&\frac{dF}{dl}=\frac{K_{d}}{\gamma \Lambda _{l}^{2-d}}\int\limits_{0}^{%
\infty }dt\,\ e^{-t}\Delta ^{\prime }(\widetilde{\mathrm{v}}t),  \label{F}
\end{eqnarray}%
where $\mathrm{v}$ is the average velocity of the interface, $\Lambda
_{l}=\Lambda _{0}e^{-l}$, $\widetilde{\mathrm{v}}=\lambda \mathrm{v}/(\gamma
\Lambda _{l}^{2})$ and $K_{d}^{-1}=2^{d-1}\pi ^{d/2}\Gamma (d/2)$. Due to
the tilt symmetry the stiffness constant $\gamma $\ does not renormalize. In
the limit $\widetilde{\mathrm{v}}\rightarrow 0$ the disorder correlator $%
\Delta (z)$ renormalizes as
\begin{equation}
\frac{d\Delta (z)}{dl}=-\frac{K_{d}}{\gamma ^{2}\Lambda _{l}^{\varepsilon }}%
\frac{d^{2}}{dz^{2}}\left[ \frac{1}{2}\Delta ^{2}(z)-\Delta (z)\Delta (0)%
\right] .  \label{delta}
\end{equation}%
The RG equations (\ref{lambda}) and (\ref{delta}) are the basis for
computation of the force-velocity characteristics in the vicinity of the
depinning transition. In following we analyze Eqs.~(\ref{lambda}) and (\ref%
{delta}) at $d=4$, i.e. for $\varepsilon =0$. Before considering the general
solution of the RG equations, we will analyze the flow equation for $\Delta
^{\prime \prime }(0)$
\begin{equation}
\frac{d\Delta ^{\prime \prime }(0)}{dl}=-\frac{3K_{4}}{\gamma ^{2}}\left[
\Delta ^{\prime \prime }(0)\right] ^{2}.  \label{detla0}
\end{equation}%
From Eq.~(\ref{detla0}) it follows that as in the case $d<d_{c}$, the second
derivative of the disorder correlator at origin $\Delta ^{\prime \prime }(0)$
diverges at the finite length $l_{c}=-\gamma ^{2}/(3K_{4}\Delta _{0}^{\prime
\prime }(0))$ for any initial condition $\Delta _{0}^{\prime \prime }(0)<0$.
Thus one obtains the Larkin length $L_{c}=\Lambda _{0}^{-1}e^{l_{c}}$ at the
upper critical dimension. The divergence of the curvature of $\Delta (z)$
implies the generation of a cusp singularity: $\Delta (z)$ becomes
non-analytical at the origin and acquires for $l>l_{c}$ a non-zero
derivative $\Delta ^{\prime }(0^+)<0$. It was shown in Ref.~\cite{nstl92}
that the cusp generated during the renormalization determines the threshold
force of the depinning transition. Therefore we expect that even at the
upper critical dimension the interface is pinned for small enough driving
force.

We now will consider the depinning transition at the upper critical
dimension. Although Eq.~(\ref{detla0}) does not have a sense beyond the
Larkin scale, nevertheless, we can still use the flow equation (\ref{delta})
for the renormalized correlator. In contrast to $d<d_{c}$ where the critical
behavior at the depinning transition is obtained from the fixed-point
solution of Eq.~(\ref{delta}) corresponding to the condition $d\Delta ^{\ast
}(z)/dl=0$, the solution of Eq.~(\ref{delta}) describing the behavior at the
depinning transition at the upper critical dimension depends explicitly on $%
l $. To find the scaling form of the function $\Delta _{l}(z)$ at $d=4$ we
look for an automodel solution of Eq.~(\ref{delta}) in the form $\Delta
_{l}(z)=K_{4}^{-1}\gamma ^{2}\phi (l)\rho (\psi (l)z)$. Note that the latter
reflects the scaling behavior at the depinning transition. Substituting this
scaling ansatz into Eq.~(\ref{delta}) we obtain the simultaneous equations
for $\phi (l)$, $\psi (l)$, and $\rho (z)$
\begin{eqnarray}
&&\phi ^{\prime }(l)=-\phi ^{2}(l)\psi ^{2}(l),\hspace{5mm}\psi ^{\prime
}(l)=-\phi (l)\psi ^{3}(l),  \label{psi} \\
&&\left( z\rho (z)\right) ^{\prime }=\left( \rho ^{2}(z)/2-\rho (z)\rho
(0)\right) ^{\prime \prime }.  \label{eqforro}
\end{eqnarray}%
Eqs.~(\ref{psi}) imply that $\psi (l)/\phi (l)=a$ is a constant which will
be determined below. This condition allows us to find $\phi
(l)=(3a^{2}l)^{-1/3}$ and $\psi (l)=(3l/a)^{-1/3}$, so that the automodel
solution of Eq.~(\ref{delta}) results in
\begin{equation}
\Delta _{l}(z)=K_{4}^{-1}\gamma ^{2}(3a^{2}l)^{-1/3}\rho (z(3l/a)^{-1/3}).
\label{ansatz}
\end{equation}%
Eq.~(\ref{ansatz}) is the pendant of the fixed-point solution of the
disorder correlator at $d<4$. One should bear in mind that the FRG equation
in this case ($d=d_{c}$) gives the exact large-scale behavior while for $%
d<d_{c}$ one must rely on the $\varepsilon $-expansion. The solution of Eq.~(%
\ref{eqforro}) with the initial condition $\rho (0)=1$, which formally
coincides with the equation for the fixed-point disorder correlator at $%
\epsilon =3$ \cite{fisher-phys-rep98}, can be written as
\begin{equation}
\rho (z)-1-\ln \rho (z)=z{^{2}}/2.  \label{solrho}
\end{equation}%
Note that $\rho (z)$ has a cusp at origin so that its behavior near $z=0$ is
given by $\rho (z)=1-\left\vert z\right\vert +\frac{1}{3}z^{2}+...$ The
constant $a$ must be defined from the initial condition for the disorder
correlator. Indeed, the flow equation (\ref{delta}) for the disorder
correlator implies that in the case of RF disorder the RF strength $%
c=\int\nolimits_{-\infty }^{+\infty }\Delta _{l}dz$ is conserved to one-loop
order \cite{nstl92} (it was shown in \cite{chauve,LeDousal-cond-mat02} that
the above integral is not conserved in the two-loop order FRG), i.e. it does
not depend on $l$. Therefore the constant $a$ in the ansatz (\ref{ansatz})
is determined by the strength $c$ of the bare disorder correlator as $%
a\approx 1.55K_{4}^{-1}\gamma ^{2}c^{-1}$, where we have used the integral $%
\int\nolimits_{-\infty }^{+\infty }\rho (z)dz\approx 1.55$. To higher orders
of FRG the non-universal constant $a$ is determined by higher moments of the
bare disorder correlator. For the bare disorder correlator being a smooth
function, the RG flow generates as in the case $d<d_{c}$ singularities on
the scale $l_{c}$, which result in the cusp of the running disorder
correlator, and therefore in the existence of the threshold force $F_{c}>0$.
Because the automodel solution (\ref{ansatz}) has the cusp on all scales,
one should use it only beyond the Larkin scale. The full solution of the
flow equation (\ref{delta}), of course, depends on the initial condition,
nevertheless, the latter is expected to approach the solution (\ref{ansatz})
in the limit $l\rightarrow \infty $. From Eq.~(\ref{ansatz}) one can
immediately derive the scaling relations for the first derivatives of the
running disorder correlator
\begin{eqnarray}
\Delta _{l}(0) &=&K_{4}^{-1}\gamma ^{2}(3a^{2}l)^{-1/3},  \label{d0} \\
\Delta _{l}^{\prime }(0^+) &=&-K_{4}^{-1}\gamma ^{2}(9al^{2})^{-1/3},
\label{d1} \\
\Delta _{l}^{\prime \prime }(0) &=&2K_{4}^{-1}\gamma ^{2}(9l)^{-1}.
\label{d2}
\end{eqnarray}

Using the above results, we will now calculate the displacement correlation
function $B(q)=\left\langle z_{q}z_{-q}\right\rangle $ that describes the
roughness of the interface at the upper critical dimension. Simple scaling
analysis shows that the correlation function satisfies the following flow
equation \cite{chitra99}
\begin{equation}
B(q)=e^{4l}B(qe^{l};\Delta _{l},F_{l}).  \label{Bq1}
\end{equation}%
In order to extract the behavior for long-wavelength correlations at the
depinning transition, $F=F_{c}$, we put $qe^{l}=\Lambda _{0}$ and expand
Eq.~(\ref{Bq1}) in powers of $\Delta $. After some algebra this yields
\begin{equation}
B(q)=\frac{\Delta _{l}(0)}{\gamma ^{2}q^{4}}=\frac{8\pi ^{2}}{(3a^{2})^{1/3}}%
\frac{1}{q^{4}\left[ \ln \Lambda _{0}/q\right] ^{1/3}},  \label{Bq2}
\end{equation}%
where in order to obtain the final expression on the right-hand side of \ (%
\ref{Bq2}) we have used Eq.~(\ref{d0}). In a direct analogy with the case $%
d<4 $ Eq.~(\ref{Bq2}) holds simultaneously in the equilibrium and at the
depinning threshold at least within the one-loop approximation \cite%
{nf93,chauve}. The Fourier transform of Eq.~(\ref{Bq2}) results in the
following real-space displacement correlation function for large distance $x$
\begin{equation}
B(x)\sim \left( \ln x\Lambda _{0}\right) ^{2/3}.  \label{result2}
\end{equation}

To obtain the force-velocity characteristics we have to integrate the flow
equations in the vicinity of $F_{c}$, i.e. in the limit of small $\widetilde{%
\mathrm{v}}$. Substituting Eq.~(\ref{d2}) into the flow equation (\ref%
{lambda}) for the friction coefficient and Eq.~(\ref{d1}) into the flow
equation for the driving force (\ref{F}) we obtain
\begin{eqnarray}
&&\frac{d\ln \lambda _{l}}{dl}=-\frac{2}{9l},  \label{eq1} \\
&&\frac{dF_{l}}{dl}=-\frac{\gamma \Lambda _{0}^{2}}{(9a)^{1/3}}%
l^{-2/3}e^{-2l}.  \label{eq2}
\end{eqnarray}%
Eqs.~(\ref{eq1}) and (\ref{eq2}) describe the renormalization of $\lambda
_{l} $ and $F_{l}$ to one-loop order beyond the Larkin scale at the upper
critical dimension. In contrast to Eqs.~(\ref{eq1}) and (\ref{eq2}) below \
the Larkin scale \ the friction coefficient $\lambda _{l}$ increases under
renormalization in accordance with Eq.~(\ref{lambda}), while the driving
force is essentially not renormalized ($dF_{l}/dl\simeq 0$). Integration of
Eqs.~(\ref{eq1}) and (\ref{eq2}) over $l$ starting from $l_{c}$ yields the
following scaling relations for the friction coefficient and the driving
force
\begin{eqnarray}
\lambda _{l} &=&\lambda _{0}(l/l_{c})^{-2/9},  \label{lambda2} \\
F_{l}-F &=&-F_{c}\simeq -0.2\Lambda _{0}^{2}(\gamma c)^{1/3},  \label{force2}
\end{eqnarray}%
where $\lambda _{0}$ is the friction coefficient on the scale $l_{c}$. In
order to obtain the renormalized friction coefficient one should express $l$
in (\ref{lambda2}) through the correlation length $L_{\mathrm{v}}$ according
to $L_{\mathrm{v}}=\Lambda _{0}^{-1}e^{l_{\mathrm{v}}}$. In $d<d_{c}$
dimensions the relation between $L_{\mathrm{v}}$ and $\mathrm{v}$ reads, $L_{%
\mathrm{v}}\sim \mathrm{v}^{-1/(z-\zeta )}$, where $z=2-(\varepsilon -\zeta
)/3$ is the dynamic exponent relating the time and the length scale [%
\onlinecite{nstl92},\onlinecite{nf93}]. Let us now derive the relation
between $L_{\mathrm{v}}$ and $\mathrm{v}$ at the upper critical dimension.
To do this we need first the relation between the time scale and the space
scale, which is derived by using the relation $t\sim \lambda
_{l}L^{2}/\gamma $ with $\lambda _{l}$ given by Eq.~(\ref{lambda2}) as $%
t\sim L^{2}(\ln L)^{-2/9}$. Following \cite{nstl92} the correlation length $%
L_{\mathrm{v}}$ can be derived by equating the systematic drift of the
interface and the height fluctuation (or equivalently from the
velocity-velocity correlation function at equal times \cite%
{chaveetal-cond-mat}), which is the square root of Eq.~(\ref{result2}), $%
\mathrm{v}t\sim (\ln L)^{1/3}$. Combining the latter with the above relation
between $t$ and $L$ gives the correlation length at the upper critical
dimension as
\begin{equation}
L_{\mathrm{v}}\sim \mathrm{v}^{-1/2}\left\vert \ln \mathrm{v}\right\vert
^{5/18}.  \label{correl-l}
\end{equation}%
The use of the relation, $F_{l_{\mathrm{v}}}=\mathrm{v}\lambda _{l_{\mathrm{v%
}}}$, which is obtained by stopping the renormalization at $l_{\mathrm{v}}$,
where $F_{l_{\mathrm{v}}}$ is given by (\ref{force2}), $\lambda _{l_{\mathrm{%
v}}}$ is given by (\ref{lambda2}) with $l_{\mathrm{v}}=\ln L_{\mathrm{v}%
}\Lambda _{0}$, and $L_{\mathrm{v}}\sim \mathrm{v}^{-1/2}$ (the logarithmic
correction to $L_{\mathrm{v}}$ in Eq.~(\ref{correl-l}) results in higher
order terms in the force-velocity characteristics) gives the implicit form
of the force-velocity characteristics in the vicinity of the depinning
transition at the upper critical dimension as
\begin{equation}
F-F_{c}\sim \frac{\mathrm{v}}{\left\vert \ln \mathrm{v}\right\vert ^{2/9}}.
\label{result1}
\end{equation}%
Within the one-loop consideration the interface velocity $\mathrm{v}$ under
the logarithm of Eq.~(\ref{result1}) can be replaced by $F-F_{c}$.
Substituting $\mathrm{v}$ from Eq.~(\ref{result1}) into Eq.~(\ref{correl-l})
we express the correlation length as function of the driving force as $L_{%
\mathrm{v}}\sim f^{-1/2}\left\vert \ln f\right\vert ^{1/6}$, where the
reduced driving force $f=F/F_{c}-1$ is introduced.

We have checked that the results (\ref{d0})-(\ref{d2}) and (\ref{lambda2})
derived here for $d=4$ are consistent with the corresponding results of Ref.
\cite{nstl92} for $d<4$, so that (\ref{v-F}) and (\ref{zeta}) tend to (\ref%
{result1}) and (\ref{result2}) for $d\rightarrow 4$, respectively.

In the recent numerical study of the depinning transition of driven
interfaces at the upper critical dimension in the random-field Ising model
(RFIM) \cite{roters-pre2002} the logarithmic corrections to the
force-velocity characteristics were chosen in the form
\begin{equation}
\left[ \mathrm{v}(f)/f\right] ^{1/\phi }\sim \left\vert \ln f\right\vert .
\label{result3}
\end{equation}%
The best fit to numerical data was obtained with $\phi =0.40\pm 0.09$.
Taking into account that the numerical determination of the logarithmic
corrections is difficult, this value is in a fair agreement with our exact
result $\phi =2/9$. The reason of the discrepancy might be due to the fact
that simulations are not carried out in the asymptotic regime, $\ln f\ll -1$%
. Indeed, the simulations in \cite{roters-pre2002} were performed for $|\ln
f|=1\div 4.5$.

In the remaining part of this Report we will discuss the contributions of
the subdominant logarithms, which appear in the two-loop order of RG. Using
the results of the work \cite{LeDousal-cond-mat02} we find that the two-loop
correction to the disorder correlator (\ref{ansatz}) at $d=4$ has the form
\begin{equation}
\Delta _{2l}(z)=K_{4}^{-1}\gamma ^{2}(9al^{2})^{-2/3}\rho
_{2}(z(3l/a)^{-1/3}),  \label{delta-II}
\end{equation}%
where $\rho _{2}(z)$ obeys the following differential equation
\begin{eqnarray}
&&\left\{ (1-\rho (z))(\rho _{2}(z)-\rho _{2}(0))+1/2[(\rho (z)-1)\rho
^{\prime }{}^{2}(z)\right. \hspace{5mm}  \notag \\
&&\mbox{}\hspace{2mm}\left. +\rho (z)]\right\} ^{\prime \prime }+z\rho
_{2}^{\prime }(z)+4\rho _{2}(z)=0,
\end{eqnarray}%
with $\rho (z)$ given by (\ref{solrho}). Following \cite{LeDousal-cond-mat02}
we impose the boundary condition $\rho _{2}(0)=0$ ($\rho _{2}(z)$ is the
counterpart of the function $y_{2}(u)$ of work \cite{LeDousal-cond-mat02},
so that this condition is required for the consistency of the results for
both $d<4$ and $d=4$). Expanding $\rho _{2}(z)$ in a Taylor series we obtain
$\rho _{2}(z)=-|z|+19/18z^{2}+...$. The leading correction to Eq.~(\ref{Bq2}%
) has the form $-\Delta^{\prime}_l(0^+)^2/(\gamma q)^4\int_k 1/k^2(k+q)^2$
\cite{LeDousal-cond-mat02}, so that the logarithmic correction to the
roughness (\ref{result2}) behaves as $(\ln x\Lambda_0)^{-1/3}$ and,
therefore, is irrelevant for large $x$. In order to take into account the
two-loop corrections to the force-velocity characteristics we need the
two-loop correction to the friction, which reads \cite{LeDousal-cond-mat02}
\begin{equation}
K_{4}^{2}/\gamma ^{4}\left( \Delta ^{\prime \prime }(0^+)^{2}+\Delta
^{\prime \prime \prime }(0^+)\Delta ^{\prime }(0^+)[3/2-\ln 2]\right) .
\label{friction2}
\end{equation}
Substituting Eq.~(\ref{delta-II}) into Eq.~(\ref{lambda}) and Eq.~(\ref%
{ansatz}) into Eq.(\ref{friction2}) we obtain the two-loop contribution to
Eq.~(\ref{eq1}) as $C/l^{2}$, where $C=-[17/2+\ln 2]/54\approx -0.17$. The
two-loop correction to the driving force (\ref{F}) results only in a shift
of the threshold force $F_{c}$, which is a non-universal quantity, so that
we will not consider it. Using the same arguments which led us to Eqs.~(\ref%
{correl-l}) and (\ref{result1}) we arrive at
\begin{eqnarray}
f \sim l_{\mathrm{v}}^{1/3}\exp(-2l_{\mathrm{v}}),\ \ \ \mathrm{v}(f)/f \sim
l_{\mathrm{v}}^{2/9}(1+C/l_{\mathrm{v}}).  \label{final}
\end{eqnarray}
Eqs.~(\ref{final}) express the force-velocity characteristics to two-loop
order of FRG by using the parametric representation in terms of the
correlation length $l_{\mathrm{v}}=\ln L_{\mathrm{v}}\Lambda _{0}$. The
latter allows us to avoid the asymptotic expansion, which leads to a
complicated expression \cite{zinn-justen}.

In order to describe the crossover to the asymptotic behavior given by Eq.~(%
\ref{result1}), one needs in addition to (\ref{final}) the corrections to
scaling. The latter are not available, since so far the only asymptotic
solution of the flow equations (\ref{lambda})-(\ref{delta}) is known. Due to
these reasons the extension of the description of the critical behavior at
the depinning transition to larger $\mathrm{v}$ is a non trivial problem.

In conclusion, we have considered the effects of quenched random field
disorder on the driven elastic interface at the upper critical dimension
close to the depinning transition. We have shown that the interface
undergoes the depinning transition at the critical driving force $F_{c}$,
and we have obtained the logarithmic corrections to the displacement
correlation function, the correlation length, and the force-velocity
characteristics. In approaching the depinning transition our one-loop
results become exact. We hope that the analytical results derived here will
be of interest for numerical studies of the depinning transition.

After this paper was finished, we learned that the more precise estimation
of $\phi $ from the simulations of RFIM \cite{lubeck-private} is in a very
good agreement with our prediction $2/9$.

A support from the Deutsche Forschungsgemeinschaft (SFB 418) is gratefully
acknowledged. We also would like to thank P. Le Doussal for a useful
discussion.

\end{document}